\documentclass{Interspeech}

\interspeechcameraready

\title{Rehearsal with Auxiliary-Informed Sampling for Audio Deepfake Detection}

\author[affiliation={1,2}]{Falih Gozi}{Febrinanto}
\author[affiliation={2}]{Kristen}{Moore}
\author[affiliation={2}]{Chandra}{Thapa}
\author[affiliation={1}]{Jiangang}{Ma}
\author[affiliation={1}]{Vidya}{Saikrishna}
\author[affiliation={3}]{Feng}{Xia}

\affiliation{Institute of Innovation, Science and Sustainability}{Federation University Australia}{Australia}
\affiliation{}{CSIRO's Data61}{Australia}
\affiliation{School of Computing Technologies}{RMIT University}{Australia}
\email{\{f.febrinanto, j.ma, v.saikrishna\}@federation.edu.au, \{kristen.moore, chandra.thapa\}@data61.csiro.au, f.xia@ieee.org}

\usepackage{stfloats} 
\usepackage{comment}
\usepackage{multirow} 
\usepackage[table]{xcolor} 
\usepackage{xcolor} 
\usepackage{hyperref} 
\definecolor{myblue}{RGB}{51, 102, 204}

\begin{document}
\maketitle
\begin{abstract}
The performance of existing audio deepfake detection frameworks degrades when confronted with new deepfake attacks. Rehearsal-based continual learning (CL), which updates models using a limited set of old data samples, helps preserve prior knowledge while incorporating new information. However, existing rehearsal techniques don't effectively capture the diversity of audio characteristics, introducing bias and increasing the risk of forgetting. To address this challenge, we propose \textbf{R}ehearsal with \textbf{A}uxiliary-\textbf{I}nformed \textbf{S}ampling (RAIS), a rehearsal-based CL approach for audio deepfake detection. RAIS employs a label generation network to produce auxiliary labels, guiding diverse sample selection for the memory buffer. Extensive experiments show RAIS outperforms state-of-the-art methods, achieving an average Equal Error Rate (EER) of 1.953\% across five experiences. The code is available at: \href{https://github.com/falihgoz/RAIS}{https://github.com/falihgoz/RAIS}.
\end{abstract}

\section{Introduction}
Deep learning has shown promising performance in audio deepfake detection~\cite{wu2020light_LCNN, liu2023leveraging_transformer, jung2022aasist_graph, tak2022wav2vec_aasist}. However, as audio deepfake generation evolves, relying on past data without adaptation leads to performance degradation over time~\cite{zhang2024remember_RWM, zhang2023you_RAWM}. Fine-tuning on new data risks catastrophic forgetting~\cite{wang2024comprehensive_CLSurveyForgetting, wang2024comprehensive_CLSurvey, salvi2025freeze}, where the model forgets previously acquired knowledge as it adapts to new information. Alternatively, retraining the model from scratch is computationally expensive and discards prior learning~\cite{wang2024comprehensive_CLSurvey, febrinanto2023graph_GLL}.

Continual learning (CL) enables models to retain knowledge while integrating new data and has been applied in computer vision, robotics, graph data, and natural language processing (NLP)~\cite{yu2025select_languageVisionCL, febrinanto2023graph_GLL}. In audio deepfake detection, CL methods such as DFWF~\cite{ma2021continual_DFWF}, RWM~\cite{zhang2024remember_RWM}, and RAWM~\cite{zhang2023you_RAWM} rely on regularization but assume no access to past data~\cite{chen2025region, chen2025continual}. However, updating the model without access to prior datasets can introduce a bias toward newly observed data~\cite{krutsylo2024inter_rehearsalBetter}.

Rehearsal-based CL overcomes this by storing and replaying past samples, with methods like Experience Replay (ER)~\cite{rolnick2019experience_ER} and ER with Asymmetric Cross-Entropy (ER-ACE)~\cite{caccia2022new_reservoir_ERACE} proving effective in reducing forgetting~\cite{hanmo2024effective_rehearsalBetter}. The challenge lies in selecting representative samples within a fixed-size memory buffer. While random sampling lacks balance, strategies such as feature centroid distance~\cite{caccia2022new_reservoir_ERACE,rolnick2019experience_reservoir}, class mean~\cite{rebuffi2017icarl_herding}, gradient-based~\cite{aljundi2019gradient_gradientSampling, aljundi2019online_MIR}, and class-balanced selection~\cite{chrysakis2020online_classBalance} rely only on primary labels (fake/real). However, audio contains diverse paralinguistic features, and ignoring them can result in a less diverse sample selection, ultimately limiting CL performance.

In this work, we propose \textbf{Rehearsal with Auxiliary-Informed Sampling (RAIS)}, which enhances stored sample quality by incorporating auxiliary labels to capture diverse audio characteristics. Since these auxiliary labels are latent, we introduce an audio auxiliary generation module that infers them via masked prediction. These labels then guide sample selection, ensuring a balanced representation of informative and diverse samples from past experiences. Our main contributions are as follows:
\begin{itemize}
    \item We propose \textbf{RAIS}, a rehearsal-based CL approach for audio deepfake detection that improves sample diversity.
    \item We develop an audio auxiliary generation module that infers auxiliary labels via masked prediction, eliminating manual labeling.
    \item We introduce an auxiliary label-informed sampling strategy that leverages the generated labels to select diverse and informative samples.
    \item Extensive experiments show RAIS outperforms state-of-the-art CL methods, achieving the lowest average EER.
\end{itemize}

\begin{figure}[!t]
  \begin{center}
    \includegraphics[width=\columnwidth]{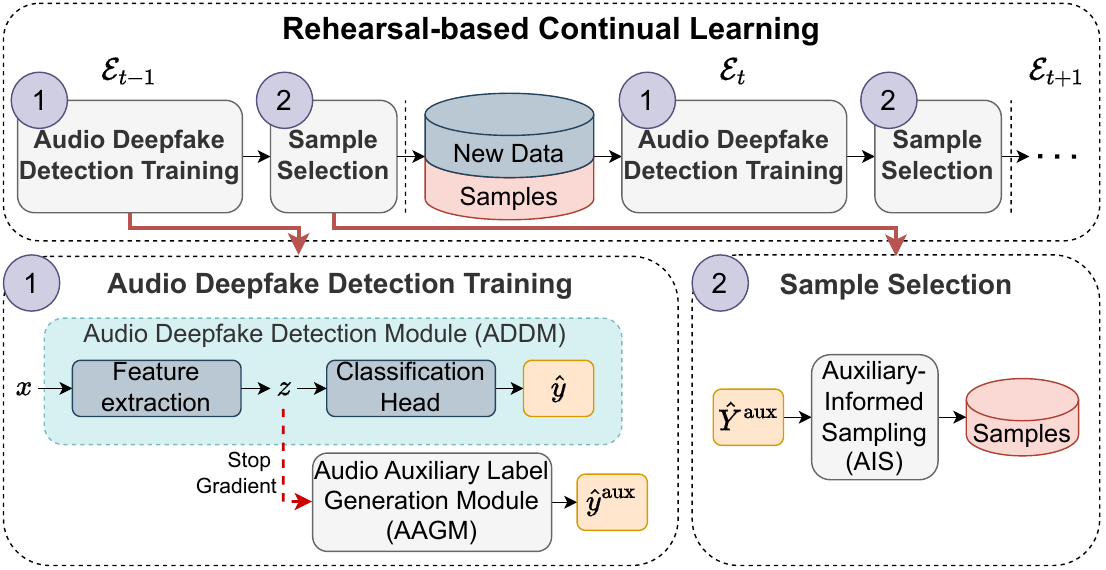}
  \end{center}
  \vspace{-10pt}
  \caption{The Proposed Framework of RAIS}
  \label{img:fram}
  \vspace{-10pt}
\end{figure}

\section{Methodology}\label{sec:proposed}
In the CL setting, training is performed across a sequence of experiences $\mathcal{E}_i$. During the initial experience $\mathcal{E}_0$, the memory buffer $\mathcal{M}$ is empty, so training relies solely on the dataset $\mathcal{D}_0$. For subsequent experiences ($i>0$), $\mathcal{D}_i$ is combined with samples from $\mathcal{M}$ to train the model while mitigating forgetting of previously learned knowledge. 

 RAIS, shown in Figure~\ref{img:fram}, enhances continual learning for audio deepfake detection through two key modules: the Audio Deepfake Detection Module (ADDM) for classification and the Audio Auxiliary Label Generation Module (AAGM) for generating auxiliary labels to guide sample selection. The auxiliary labels help mitigate catastrophic forgetting by ensuring diverse, informative samples are retained.

\subsection{Audio Deepfake Detection Training}
RAIS employs two key modules: ADDM, which classifies audio as fake or bona fide, and AAGM, which generates auxiliary labels to improve sample selection. These two modules are jointly trained to enhance learning and knowledge retention.

\subsubsection{Audio Deepfake Detection Module (ADDM)} 
ADDM consists of two components: a feature extractor $g$ that encodes input audio into a latent representation and a classification head $c$  that maps the latent representation to logits. Given an input audio signal $x \in \mathbb{R}^T$, the ADDM processes it as $\text{SoftMax}(c(g(x)))$, where $g(x)$ produces a latent representation $z \in \mathbb{R}^D$, and $c$ maps $z$ to logits. These logits are then converted into a probability distribution $p = (p_0, p_1)$ using the SoftMax function, where $p_0$ represents the probability of fake audio ($y = 0$) and $p_1$ represents the probability of bona fide audio ($y = 1$). The final classification decision is $\hat{y} = \arg\max(p)$. The ADDM is optimized using the cross-entropy loss 
$\mathcal{L}_{\text{ADDM}}$ 
which encourages the model to correctly classify audio as either fake or bona fide.

\subsubsection{Audio Auxiliary Label Generation Module (AAGM)}

While primary labels (fake/bona fide) provide essential supervision, they fail to capture the rich paralinguistic characteristics inherent in audio. To address this, we introduce AAGM, which automatically generates auxiliary labels to guide learning. AAGM is inspired by meta-auxiliary learning~\cite{liu2019self_auxiliaryTask}, but differs in key ways: \\
1. Decoupled Optimization: Unlike conventional meta-auxiliary learning, which jointly optimizes both the main task and label generation, AAGM prevents \textit{conflicting gradients}~\cite{ding2023mitigating_conflictingGradients} that could degrade primary task performance. This is achieved using a \textbf{stop-gradient} to isolate AAGM updates. \\
2. Independent Training: Instead of a multi-task objective, we train AAGM separately using a masked prediction objective~\cite{baevski2023efficient_maskedObjective}, ensuring the auxiliary network remains independent of ADDM.

\begin{figure}[!ht]
\centering
\includegraphics[width=\columnwidth]{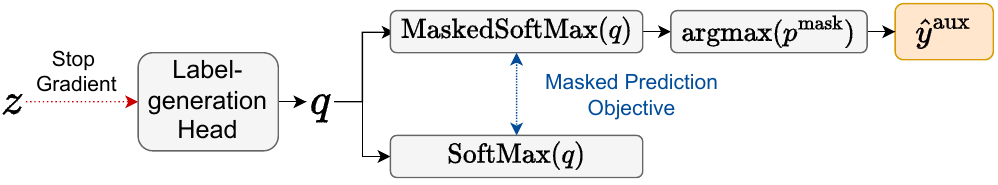}
\caption{Audio Auxiliary Label Generation Module (AAGM)}
\label{img:aagm}
\vspace{-10pt}
\end{figure}

Figure~\ref{img:aagm} illustrates AAGM's design. It begins with a detached latent representation $z$ from the feature extractor. The label-generation head $h$ produces a vector $q = h(z) \in \mathbb{R}^K$, where $K$ is the number of possible auxiliary labels. For each input, AAGM categorizes it into one of these $K$ labels. AAGM then generates two multi-class probability vectors through separate branches:

\textit{1. Masked branch (Auxiliary-Specific SoftMax):} To enforce distinct auxiliary labels for fake and bona fide samples, we apply MaskedSoftMax~\cite{liu2019self_auxiliaryTask}. The first $K/2$ labels are assigned to fake audio, while the remaining $K/2$ labels correspond to bona fide, ensuring category-specific labels. For each sample, a binary mask vector $M$ is constructed based on its ground-truth label $y$. Specifically, if the sample is fake ($y=0$), the mask activates only the first $K/2$ positions, setting the remaining positions to zero. If the sample is bona fide ($y=1$), the mask activates only the last $K/2$ positions, leaving the others at zero. This mask ensures that each sample is only assigned to valid auxiliary labels within its respective class. The masked probability vector $p^{\text{mask}}$ is then computed as:

\begin{equation}
    p^{\text{mask}} = \frac{\exp(q) \odot M}{\sum \left(\exp(q) \odot M\right)}.
\end{equation} 

\textit{2. Unmasked branch (Standard SoftMax):} The second branch calculates a standard \textit{SoftMax} probability vector  $p^{\textup{unmask}}$.

AAGM is trained with a combination of two losses: (i) MSE Loss: which aligns the masked and unmasked probability vectors, and (ii) Diversity Loss (KL Divergence): which prevents trivial solutions by ensuring a uniform distribution of generated labels:
\begin{equation}
\mathcal{L}_{\text{AAGM}} = \underbrace{\Vert p^{\text{mask}} - p^{\text{unmask}} \Vert^2}_{\text{MSE Loss}} + \underbrace{\text{KL}\Big(\bar{p}^{\text{mask}} \,\Vert\, \frac{1}{K}\mathbf{1}\Big)}_{\text{Diversity Loss}},
\end{equation}
where $\bar{p}^{\text{mask}}$ is the average of masked probabilities over the batch, and $\frac{1}{K}\mathbf{1}$ is the uniform distribution over $K$ auxiliary classes. 
The final auxiliary label is determined as 
    $\hat{y}^\text{aux} = \arg\max(p^{\text{mask}})$.

\textbf{Overall Objective.} The final training loss is $\mathcal{L} = \mathcal{L}_{\text{ADDM}} + \mathcal{L}_{\text{AAGM}}$ with stop-gradient decoupling, ensuring that AAGM does not interfere with the primary audio deepfake detection ADDM task training.

\subsection{Sample Selection via Auxiliary-Informed Sampling}
Traditional sampling methods rely solely on primary labels (fake or bona fide), which can lead to poor sample diversity. We propose Auxiliary-Informed Sampling (AIS), a novel strategy that leverages auxiliary labels to ensure diversity and informativeness in selected samples.

AIS maintains a memory buffer $\mathcal{M}$ composed of segments 
$\{\mathcal{G}_1, \mathcal{G}_2, \dots, \mathcal{G}_i\}$, where each $\mathcal{G}_k$ corresponds to a past experience $\mathcal{E}_k$. For each new experience $\mathcal{E}_i$, the allocation size is set as $L = \frac{|\mathcal{M}|}{i+1}$. After training on $\mathcal{E}_i$, $L$ samples are selected and added to $\mathcal{M}$. Since the memory buffer has a fixed size, each past memory segment $\mathcal{G}_k$ is updated to retain only the top $L$ most representative samples, ensuring that past experiences remain well-represented.

The AIS strategy first partitions the dataset into two primary categories: fake samples, $\mathcal{D}_{i, \hat{y}_{\text{aux}} \in [0, \frac{K}{2} - 1]}$, 
and bona fide samples, $\mathcal{D}_{i, \hat{y}_{\text{aux}} \in [\frac{K}{2}, K - 1]}$. Within each category, samples are further divided into groups based on their auxiliary labels. For example, in the bona fide category, the groups are indexed from $K/2$ to $K-1$. Each group is then sorted in descending order based on an importance score $s$, defined as:
\begin{equation}
\label{eq:score}
s = \frac{1}{2} \left(p_{\hat{y}} + p_{\hat{y}^{\text{aux}}}^{\text{mask}} \right),
\end{equation} 
where $p_{\hat{y}}$ is the classification confidence from ADMM, and $p_{\hat{y}_{\text{aux}}}^{\text{mask}}$ is the confidence score from AAGM. The scoring mechanism prioritizes samples with higher overall confidence, ensuring that the most reliable and informative samples are retained.

To maintain class balance, AIS introduces a ratio $r$, selecting $L \times r$ fake samples and $L \times (1 - r)$ bona fide samples. AIS then performs stratified selection using a round-robin approach across the auxiliary label groups within each category, ensuring each auxiliary label is represented within its category in $\mathcal{M}$. If a group runs out of samples, the process continues with the remaining groups until the required number of samples is reached. Finally, the selected samples from both categories are merged and re-sorted in descending order by the importance score $s$, forming a new memory segment $\mathcal{G}_i$. This segment is then integrated into the memory buffer $\mathcal{M}$, ensuring past experiences remain well-represented.

\section{Experiments}

\begin{table*}[!ht]
\label{tb:datasetStat}
\centering
\caption{Dataset statistics across different experiences in the CL setting.}
\label{tab:stats}
\resizebox{\textwidth}{!}{
\begin{tabular}{c|ccc|ccc|ccc|ccc|ccc}
\hline
\multirow{2}{*}{$\mathcal{E}$} 
& \multicolumn{3}{c|}{\begin{tabular}[c]{@{}c@{}}\textbf{ASVSpoof 2019 LA} \\ $\mathcal{E}_0$ (English)\end{tabular}} 
& \multicolumn{3}{c|}{\begin{tabular}[c]{@{}c@{}}\textbf{VCC 2020} \\ $\mathcal{E}_1$ (Multi-language)\end{tabular}} 
& \multicolumn{3}{c|}{\begin{tabular}[c]{@{}c@{}}\textbf{InTheWild} \\ $\mathcal{E}_2$ (English)\end{tabular}} 
& \multicolumn{3}{c|}{\begin{tabular}[c]{@{}c@{}}\textbf{CFAD} \\ $\mathcal{E}_3$ (Chinese)\end{tabular}} 
& \multicolumn{3}{c}{\begin{tabular}[c]{@{}c@{}}\textbf{OpenAI-LJSpeech} \\ $\mathcal{E}_4$ (English)\end{tabular}} \\ \cline{2-16} 
 & Fake & Bona fide & Total & Fake & Bona fide & Total & Fake & Bona fide & Total & Fake & Bona fide & Total & Fake & Bona fide & Total \\ \hline
Train & 22,800 & 2,580 & 25,380 & 2,920 & 805 & 3,725 & 5,908 & 9,981 & 15,889 & 25,600 & 12,800 & 38,400 & 6,550 & 6,550 & 13,100 \\
Dev & 22,296 & 2,548 & 24,844 & 1,460 & 402 & 1,862 & 2,954 & 4,991 & 7,945 & 9,600 & 4,800 & 14,400 & 3,275 & 3,275 & 6,550 \\
Eval & 63,882 & 7,355 & 71,237 & 1,460 & 403 & 1,863 & 2,954 & 4,991 & 7,945 & 42,000 & 21,000 & 63,000 & 3,275 & 3,275 & 6,550 \\ \hline
\end{tabular}
}
\vspace{-10pt}
\end{table*}

\subsection{Datasets and Experimental Settings}

\textbf{Datasets}. We evaluate our method and the baseline methods in a CL setting for audio deepfake detection across five experiences. The initial experience, $\mathcal{E}_0$, uses ASVspoof 2019 LA~\cite{todisco2019asvspoof_asvspoof2019} with its original splits (training, development, and evaluation). Experiences $\mathcal{E}_1$ and $\mathcal{E}_2$ consist of the VCC 2020~\cite{todisco2019asvspoof_asvspoof2019} and InTheWild~\cite{muller2022does_itw} datasets, each split into 25\% development, 25\% evaluation, and 50\% training. Experience $\mathcal{E}_3$, is the CFAD~\cite{ma2024cfad} dataset using its provided splits (training, development, and evaluation, combining seen and unseen tests). To incorporate more advanced speech generation tools, the final experience is generated with the OpenAI TTS API\footnote{\href{https://platform.openai.com/docs/guides/text-to-speech}{https://platform.openai.com/docs/guides/text-to-speech}} using scripts from LJSpeech\footnote{\href{https://keithito.com/LJ-Speech-Dataset/}{https://keithito.com/LJ-Speech-Dataset/}}. Each transcript was synthesized with a random OpenAI TTS voice (\textit{alloy}, \textit{echo}, \textit{fable}, \textit{onyx}, \textit{nova}, \textit{shimmer}) and model type (\textit{tts-1} or \textit{tts-1-hd}). Fake samples were generated by OpenAI TTS, while bona fide samples were from LJSpeech; both were split with the same proportions as VCC 2020 and InTheWild. Dataset statistics are provided in Table~\ref{tab:stats}.

\noindent \textbf{Experimental Settings}. Audio clips were standardized to 4 seconds. We used Wav2Vec2~\cite{baevski2020wav2vec} (wav2vec2-xls-r-300m) as the front-end to convert raw audio into a 2D matrix, then passed it to AASIST~\cite{jung2022aasist_aasist,tak2022wav2vec_aasist} for feature extraction. The model was trained with a dropout rate of 0.1, a batch size of 64, and the Adam optimizer with a learning rate of 0.00001. Training was conducted for 10 epochs with early stopping. The classification head $c(\cdot)$ and the label-generation head $h(\cdot)$ consist of two linear layers with hidden dimensions of 80 and 32. The auxiliary label size was set to $K=90$, with parameter sensitivity analysis provided in Section~\ref{sec:abla}. The fake/bona fide ratio was set to $r=0.8$. For evaluation, we used the equal error rate (EER) and the average EER across all experiences. Each baseline was run three times with different seeds, reporting the mean and standard deviation.

\begin{table*}[!ht]
\centering
\caption{EER (\%) comparison of CL methods trained sequentially (\(\mathcal{E}_0 \rightarrow \mathcal{E}_1 \rightarrow \mathcal{E}_2 \rightarrow \mathcal{E}_3 \rightarrow \mathcal{E}_4\)) and evaluated on all test sets after the final stage. Lower values indicate better performance (\(\downarrow\)). \textbf{Bold} marks the best result, \underline{underline} denotes the second-best, and \textcolor{blue!80}{Blue} highlights non-CL methods.}
\label{tb:baselines}
\resizebox{\textwidth}{!}{
\begin{tabular}{cccllllll}
\hline
\multicolumn{1}{c|}{\textbf{Buffer Size}} & \multicolumn{1}{c|}{\textbf{Method}} & \multicolumn{1}{c|}{\textbf{Sampling}} & \multicolumn{1}{c}{\textbf{$\mathcal{E}_0$}} & \multicolumn{1}{c}{\textbf{$\mathcal{E}_1$}} & \multicolumn{1}{c}{\textbf{$\mathcal{E}_2$}} & \multicolumn{1}{c}{\textbf{$\mathcal{E}_3$}} & \multicolumn{1}{c|}{\textbf{$\mathcal{E}_4$}} & \textbf{Avg EER.} \\ \hline
\multicolumn{1}{c|}{-} & \multicolumn{1}{c|}{\textcolor{blue!80}{Trained on $\mathcal{E}_0$}} & \multicolumn{1}{c|}{-} & \textcolor{blue!80}{7.836 $\pm$ 5.820} & \textcolor{blue!80}{2.481 $\pm$ 0.993} & \textcolor{blue!80}{8.041 $\pm$ 0.133} & \textcolor{blue!80}{10.170 $\pm$ 1.195} & \multicolumn{1}{l|}{\textcolor{blue!80}{20.193 $\pm$ 11.098}} & \textcolor{blue!80}{9.744 $\pm$ 3.260} \\
\multicolumn{1}{c|}{-} & \multicolumn{1}{c|}{\textcolor{blue!80}{Trained on all}} & \multicolumn{1}{c|}{-} & \textcolor{blue!80}{0.553 $\pm$ 0.126} & \textcolor{blue!80}{0.000 $\pm$ 0.000} & \textcolor{blue!80}{0.294 $\pm$ 0.023} & \textcolor{blue!80}{8.013 $\pm$ 0.203} & \multicolumn{1}{l|}{\textcolor{blue!80}{0.000 $\pm$ 0.000}} & \textcolor{blue!80}{1.772 $\pm$ 0.030} \\
\multicolumn{1}{c|}{-} & \multicolumn{1}{c|}{\textcolor{blue!80}{Fine-tune}} & \multicolumn{1}{c|}{-} & \textcolor{blue!80}{1.446 $\pm$ 0.432} & \textcolor{blue!80}{6.865 $\pm$ 4.155} & \textcolor{blue!80}{4.855 $\pm$ 1.387} & \textcolor{blue!80}{9.802 $\pm$ 1.127} & \multicolumn{1}{l|}{\textcolor{blue!80}{0.000 $\pm$ 0.000}} & \textcolor{blue!80}{4.594 $\pm$ 1.071} \\ \hline
\multicolumn{1}{c|}{-} & \multicolumn{1}{c|}{EWC} & \multicolumn{1}{c|}{-} & $1.894 \pm 0.605$ & $7.858 \pm 4.228$ & $6.044 \pm 3.486$ & $11.241 \pm 2.215$ & \multicolumn{1}{l|}{\textbf{0.000 $\pm$ 0.000}} & $5.408 \pm 1.950$ \\
\multicolumn{1}{c|}{-} & \multicolumn{1}{c|}{LwF} & \multicolumn{1}{c|}{-} & $1.523 \pm 0.501$ & $2.647 \pm 1.249$ & \textbf{0.521} $\pm$ 0.223 & $10.052 \pm 0.731$ & \multicolumn{1}{l|}{$0.275 \pm 0.153$} & 3.004 $\pm$ 0.552 \\
\multicolumn{1}{c|}{-} & \multicolumn{1}{c|}{OWM} & \multicolumn{1}{c|}{-} & $1.550 \pm 0.419$ & $4.301 \pm 2.940$ & $6.438 \pm 2.002$ & $11.343 \pm 2.247$ & \multicolumn{1}{l|}{\textbf{0.000 $\pm$ 0.000}} & $4.726 \pm 0.376$ \\
\multicolumn{1}{c|}{-} & \multicolumn{1}{c|}{RAWM} & \multicolumn{1}{c|}{-} & $1.627 \pm 1.206$ & $5.376 \pm 6.090$ & $4.234 \pm 1.898$ & $10.451 \pm 1.582$ & \multicolumn{1}{l|}{\textbf{0.000 $\pm$ 0.000}} & $4.338 \pm 2.138$ \\
\multicolumn{1}{c|}{-} & \multicolumn{1}{c|}{RWM} & \multicolumn{1}{c|}{-} & 1.428 $\pm$ 0.987 & $4.549 \pm 3.152$ & $4.147 \pm 0.626$ & $11.211 \pm 0.186$ & \multicolumn{1}{l|}{\textbf{0.000 $\pm$ 0.000}} & $4.267 \pm 0.779$ \\ \hline
\multicolumn{1}{c|}{256} & \multicolumn{1}{c|}{ER} & \multicolumn{1}{c|}{MIR} & 0.970 $\pm$ 0.450 & $3.060 \pm 1.249$ & $1.743 \pm 0.106$ & $9.029 \pm 0.998$ & \multicolumn{1}{l|}{$0.020 \pm 0.018$} & $2.964 \pm 0.412$ \\
\multicolumn{1}{c|}{256} & \multicolumn{1}{c|}{ER} & \multicolumn{1}{c|}{Class-balanced} & 1.215 $\pm$ 0.169 & 1.406 $\pm$ 0.517 & 1.122 $\pm$ 0.459 & 8.151 $\pm$ 0.172 & \multicolumn{1}{l|}{\underline{0.010 $\pm$ 0.018}} & 2.381 $\pm$ 0.180 \\
\multicolumn{1}{c|}{256} & \multicolumn{1}{c|}{ER} & \multicolumn{1}{c|}{Reservoir} & \textbf{0.657 $\pm$ 0.172} & $1.241 \pm 0.248$ & $1.837 \pm 0.076$ & $8.171 \pm 0.215$ & \multicolumn{1}{l|}{\textbf{0.000 $\pm$ 0.000}} & $2.381 \pm 0.033$ \\
\multicolumn{1}{c|}{256} & \multicolumn{1}{c|}{ER} & \multicolumn{1}{c|}{Herding} & $2.479 \pm 1.641$ & \underline{0.331 $\pm$ 0.143} & $1.282 \pm 0.278$ & $8.351 \pm 0.060$ & \multicolumn{1}{l|}{\textbf{0.000 $\pm$ 0.000}} & $2.489 \pm 0.292$ \\
\multicolumn{1}{c|}{256} & \multicolumn{1}{c|}{ER-ACE} & \multicolumn{1}{c|}{Class-balanced} & 0.879 $\pm$ 0.348 & 1.241 $\pm$ 0.430 & 1.563 $\pm$ 0.386 & 8.197 $\pm$ 0.110 & \multicolumn{1}{l|}{\underline{0.010 $\pm$ 0.018}} & 2.378 $\pm$ 0.131 \\
\multicolumn{1}{c|}{256} & \multicolumn{1}{c|}{ER-ACE} & \multicolumn{1}{c|}{Reservoir} & $1.169 \pm 0.575$ & $1.654 \pm 0.379$ & $2.017 \pm 0.418$ & $8.119 \pm 0.765$ & \multicolumn{1}{l|}{$0.020 \pm 0.035$} & $2.596 \pm 0.126$ \\
\multicolumn{1}{c|}{256} & \multicolumn{1}{c|}{ER-ACE} & \multicolumn{1}{c|}{Herding} & $1.106 \pm 0.417$ & $1.489 \pm 1.082$ & $1.710 \pm 0.313$ & \textbf{7.930 $\pm$ 0.357} & \multicolumn{1}{l|}{\underline{0.010 $\pm$ 0.018}} & $2.449 \pm 0.240$ \\
\multicolumn{1}{c|}{256} & \multicolumn{1}{c|}{\textbf{RAIS}} & \multicolumn{1}{c|}{\textbf{AIS}} & $0.847 \pm 0.143$ & 0.248 $\pm$ 0.248 & 2.070 $\pm$ 0.182 & \underline{7.971 $\pm$ 0.349} & \multicolumn{1}{l|}{\textbf{0.000 $\pm$ 0.000}} & 2.230 $\pm$ 0.095 \\ \hline
\multicolumn{1}{c|}{512} & \multicolumn{1}{c|}{ER} & \multicolumn{1}{c|}{MIR} & $1.541 \pm 0.594$ & $2.564 \pm 1.433$ & $1.316 \pm 0.281$ & $8.492 \pm 0.454$ & \multicolumn{1}{l|}{\underline{0.010 $\pm$ 0.018}} & $2.785 \pm 0.343$ \\
\multicolumn{1}{c|}{512} & \multicolumn{1}{c|}{ER} & \multicolumn{1}{c|}{Class-balanced} & 0.979 $\pm$ 0.756 & 0.910 $\pm$ 0.379 & 0.928 $\pm$ 0.292 & 8.365 $\pm$ 0.138 & \multicolumn{1}{l|}{\textbf{0.000 $\pm$ 0.000}} & 2.236 $\pm$ 0.259 \\
\multicolumn{1}{c|}{512} & \multicolumn{1}{c|}{ER} & \multicolumn{1}{c|}{Reservoir} & $1.260 \pm 0.265$ & $0.744 \pm 0.430$ & $1.002 \pm 0.122$ & $8.108 \pm 0.307$ & \multicolumn{1}{l|}{\underline{0.010 $\pm$ 0.018}} & $2.225 \pm 0.094$ \\
\multicolumn{1}{c|}{512} & \multicolumn{1}{c|}{ER} & \multicolumn{1}{c|}{Herding} & 1.187 $\pm$ 0.532 & \underline{0.331 $\pm$ 0.143} & 0.922 $\pm$ 0.209 & 8.043 $\pm$ 0.203 & \multicolumn{1}{l|}{\textbf{0.000 $\pm$ 0.000}} & \underline{2.097 $\pm$ 0.067} \\
\multicolumn{1}{c|}{512} & \multicolumn{1}{c|}{ER-ACE} & \multicolumn{1}{c|}{Class-balanced} & 1.296 $\pm$ 0.263 & 2.647 $\pm$ 0.758 & 0.848 $\pm$ 0.117 & 8.179 $\pm$ 0.200 & \multicolumn{1}{l|}{\underline{0.010 $\pm$ 0.018}} & 2.596 $\pm$ 0.198 \\
\multicolumn{1}{c|}{512} & \multicolumn{1}{c|}{ER-ACE} & \multicolumn{1}{c|}{Reservoir} & $0.934 \pm 0.188$ & $2.233 \pm 0.657$ & $1.289 \pm 0.189$ & $8.348 \pm 0.232$ & \multicolumn{1}{l|}{\underline{0.010 $\pm$ 0.018}} & $2.563 \pm 0.222$ \\
\multicolumn{1}{c|}{512} & \multicolumn{1}{c|}{ER-ACE} & \multicolumn{1}{c|}{Herding} & $0.748 \pm 0.286$ & $1.820 \pm 0.798$ & $1.423 \pm 0.434$ & 8.014 $\pm$ 0.053 & \multicolumn{1}{l|}{\underline{0.010 $\pm$ 0.018}} & $2.403 \pm 0.182$ \\
\multicolumn{1}{c|}{512} & \multicolumn{1}{c|}{\textbf{RAIS}} & \multicolumn{1}{c|}{\textbf{AIS}} & \underline{0.666
 $\pm$ 0.187} & \textbf{0.083 $\pm$ 0.143} & \underline{0.821 $\pm$ 0.151} & 8.197 $\pm$ 0.184 & \multicolumn{1}{l|}{{\textbf{0.000 $\pm$ 0.000}}} & \textbf{1.953 $\pm$ 0.106} \\ \hline
\end{tabular}
 }
\end{table*}

\subsection{Experimental Results}\label{sec:exp_result}
\noindent \textbf{Baselines}. We compare our method against several baselines. The naive baselines include “Trained on $\mathcal{E}_0$,” where the model is trained only on the first dataset and evaluated on all experiences; “Trained on all,” where all datasets are merged for the best possible performance; and “Fine-tune,” where the model is updated on each new experience without strategies to mitigate forgetting. For regularization-based continual learning methods with no memory buffer, we include EWC~\cite{kirkpatrick2017overcoming_EWC}, LwF~\cite{li2017learning_LWF}, and OWM~\cite{zeng2019continual_OWM}, as well as 2 audio deepfake-specific methods: RAWM~\cite{zhang2023you_RAWM} and RWM~\cite{zhang2024remember_RWM}. Additionally, we evaluate rehearsal-based strategies with buffer sizes of 256 and 512, including ER~\cite{rolnick2019experience_ER} and ER-ACE~\cite{caccia2022new_reservoir_ERACE}, where we use different sample selection strategies, including reservoir~\cite{caccia2022new_reservoir_ERACE}, herding~\cite{rebuffi2017icarl_herding}, and class-balanced~\cite{chrysakis2020online_classBalance} for ER and ER-ACE, and additionally MIR~\cite{aljundi2019online_MIR} for ER only. Lastly, we excluded the gradient-based sampling method GSS~\cite{aljundi2019gradient_gradientSampling} as it led to \textit{out-of-memory} issues, making it impractical for high-dimensional audio data. 

\noindent \textbf{Results and Analysis}. Table~\ref{tb:baselines} presents the results, showing that RAIS with a 512-sample buffer achieves an average EER of 1.953\%, closely matching the best possible performance of the Train-on-all approach. Overall, CL strategies demonstrate competitive performance relative to Train-on-all while being significantly more efficient, eliminating the need for retraining from scratch with each new experience. In contrast, naive methods such as training exclusively on the initial experience $\mathcal{E}_0$ quickly become obsolete as new experiences emerge, while fine-tuning alone fails to retain knowledge from earlier tasks. Replay-based strategies consistently outperform regularization-based approaches (e.g., EWC, LwF, RAWM, and RWM), which lack memory buffers and cannot leverage past data. However, these methods are not directly comparable, as rehearsal-based methods, regularization-based CL retains knowledge by applying constraints to model updates instead of revisiting past data. Compared to other rehearsal strategies such as Experience Replay (ER) and ER with Asymmetric Cross-Entropy (ER-ACE), RAIS further improves performance by introducing an additional network branch that generates diverse auxiliary labels, leading to better sample selection and a more representative memory buffer based on meaningful and diverse audio characteristics.

\begin{table}[!ht]
\centering
\caption{EER (\%) comparisons of RAIS and its ablated variants, where lower is better ($\downarrow$). \textbf{Bold} denotes the best result.}
\label{tb:ablation}
\resizebox{\columnwidth}{!}{
\begin{tabular}{l|ccccc}
\hline
\textbf{Method} & \textbf{$\mathcal{E}_0$} & \textbf{$\mathcal{E}_1$} & \textbf{$\mathcal{E}_2$} & \textbf{$\mathcal{E}_3$} & \textbf{$\mathcal{E}_4$} \\ \hline
(-) AAGM & 3.89 $\pm$ 2.35 & 1.90 $\pm$ 1.65 & 2.61 $\pm$ 1.99 & 8.74 $\pm$ 0.09 & 0.01 $\pm$ 0.02 \\
(-) AIS & 3.56 $\pm$ 1.72 & 3.08 $\pm$ 3.37 & 3.01 $\pm$ 1.51 & 8.81 $\pm$ 0.57 & 0.01 $\pm$ 0.01 \\
(-) DL & 1.78 $\pm$ 0.65 & 0.45 $\pm$ 0.32 & 2.28 $\pm$ 0.61 & \textbf{8.16 $\pm$ 0.14} & 0.01 $\pm$ 0.01 \\ \hline
\textbf{RAIS} & \textbf{0.67 $\pm$ 0.19} & \textbf{0.08 $\pm$ 0.14} & \textbf{0.82 $\pm$ 0.15} & 8.20 $\pm$ 0.18 & \textbf{0.00 $\pm$ 0.00} \\ 
\hline
\end{tabular}
}
\end{table}

\subsection{Ablation Studies and Parameter Sensitivity}\label{sec:abla}

\textbf{Ablation Studies}. We performed ablation studies using a 512-sample buffer, removing key components one at a time. Three configurations were evaluated. First, RAIS without the Audio Auxiliary Label Generation Module (AAGM), denoted (-) AAGM, which relies solely on main task labels and their confidence scores for sample selection. Second, RAIS without Auxiliary-Informed Selection (AIS), denoted (-) AIS, where sample selection is based only on the highest score (Equation~\ref{eq:score}) without diversifying across auxiliary label groups. Third, RAIS without Diversity Loss, denoted as (-) DL, removes the diversity loss in $\mathcal{L}_{\text{AAGM}}$. Table~\ref{tb:ablation} shows that the complete RAIS consistently outperforms its ablated variants.

\begin{figure}[!ht]
\centering
\includegraphics[width=\columnwidth]{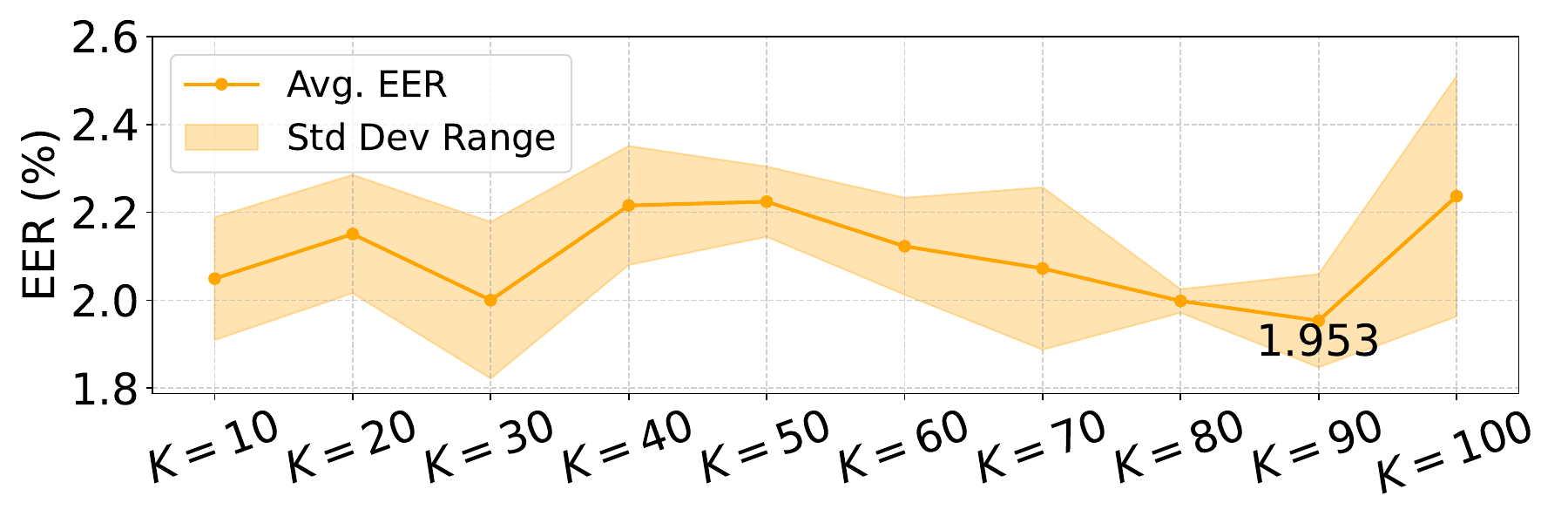}
\caption{Average EER (\%) across different experiences for varying auxiliary label size $K$.}
\label{img:sens}
\vspace{-10pt}
\end{figure}

\noindent \textbf{Parameter Sensitivity}. We assessed the sensitivity of the auxiliary label size $K$, which determines the number of possible labels predicted by AAGM. We analyzed $K$ values of 10, 20, 30, 40, 50, 60, 70, 90, and 100. As shown in Figure~\ref{img:sens}, $K=90$ achieved the best performance with the lowest average EER.

\begin{table}[]
\caption{Forgetting Rate Comparisons on $\mathcal{E}_1$. CL methods are trained sequentially ($\mathcal{E}_1 \rightarrow \mathcal{E}_2$). Lower values indicate better performance ($\downarrow$). \textbf{Bold} denotes the best result, and \textcolor{blue!80}{Blue} denotes non-CL methods.}
\label{tb:forgetting}
\resizebox{\columnwidth}{!}{
\begin{tabular}{c|c|c|cc|c}
\hline
\textbf{Buffer Size} & \textbf{Method} & \textbf{Sampling} & \textbf{$\mathcal{E}_1$} & \textbf{$\mathcal{E}_2$} & \textbf{$F_1$} \\ \hline
- & \textcolor{blue!80}{Trained on $\mathcal{E}_1$} & - & \textcolor{blue!80}{0.00 $\pm$ 0.00} & \textcolor{blue!80}{10.75 $\pm$ 3.88} & - \\
- & \textcolor{blue!80}{Trained on all} & - & \textcolor{blue!80}{0.08 $\pm$ 0.14} & \textcolor{blue!80}{0.36 $\pm$ 0.17} & - \\
- & \textcolor{blue!80}{Fine-tune} & - & \textcolor{blue!80}{6.12 $\pm$ 1.52} & \textcolor{blue!80}{0.31 $\pm$ 0.02} & + (\textcolor{blue!80}{6.12 $\pm$ 1.52}) \\ \hline
512 & ER & Herding & 0.33 $\pm$ 0.38 & 0.34 $\pm$ 0.08 & + (0.33 $\pm$ 0.38)  \\
512 & \textbf{RAIS} & \textbf{AIS} & 0.17 $\pm$ 0.14 & 0.29 $\pm$ 0.07 & + (\textbf{0.17 $\pm$ 0.14}) \\ \hline
\end{tabular}}
\vspace{-10pt}
\end{table}

\subsection{Discussion}

\noindent \textbf{Forgetting in Audio Deepfake Detection}. 
Surprisingly, Table~\ref{tb:baselines} shows no forgetting of $\mathcal{E}_0$ knowledge when comparing ``Trained on $\mathcal{E}_0$" with ``Fine-Tune." This may be due to the presence of unseen deepfake attacks in the ASVspoof 2019 LA ($\mathcal{E}_0$) evaluation set, which were not part of the training data (similarly for $\mathcal{E}_3$). However, some of these unseen deepfake generators—or generators with similar fingerprints—may appear in the training sets of later experiences. Fine-tuning on new experiences can therefore, in some cases, improve performance by exposing the model to previously unseen patterns. This is reflected in the EER drop for $\mathcal{E}_0$ from 7.836\% (trained solely on $\mathcal{E}_0$) to 1.446\% after sequential fine-tuning.

To directly assess forgetting, we designed an experiment where the evaluation set only includes attack types that are present in the training set, ensuring similar generator fingerprints in the train, development, and test splits. For this, we used datasets from $\mathcal{E}_1$ (VCC 2020) and $\mathcal{E}_2$ (InTheWild). Table~\ref{tb:forgetting} reports the forgetting rate $F_1$ for $\mathcal{E}_1$, defined as the performance drop after sequentially learning $\mathcal{E}_1 \rightarrow \mathcal{E}_2$ compared to training on $\mathcal{E}_1$ alone. Notably, RAIS consistently achieves the lowest forgetting rate (0.17\%), outperforming the second-best baseline ER with Herding, demonstrating superior mitigation of forgetting in CL scenarios.

\noindent \textbf{Limitations and Future Directions}. 
RAIS maintains CL performance with a 512-sample memory buffer ($\sim$ 50 MB), offering high efficiency compared to full dataset storage ($>$50 GB). Future work should explore scalability for larger CL scenarios. Secondly, while RAIS prioritizes fake samples to reduce privacy risks, storing genuine audio may still pose concerns. Investigating privacy-preserving techniques such as differential privacy could help mitigate these risks. Additionally, the interpretability of auxiliary labels remains an open question. Future research should examine what these labels capture, how they evolve over time, and whether they provide deeper insights into deepfake detection.

\section{Conclusion}\label{sec:conclu}
We introduced Rehearsal with Auxiliary-informed Sampling (RAIS), a CL approach for audio deepfake detection. RAIS improves sample diversity in the memory buffer by automatically generating auxiliary labels, capturing diverse audio characteristics without the need for manual labeling. These labels guide sample selection, ensuring a balanced representation of audio features. Extensive experiments show that RAIS outperforms state-of-the-art CL and experience replay methods across five experiences, achieving the lowest average EER.

\newpage
\bibliographystyle{IEEEtran}
\bibliography{mybib}

\begin{thebibliography}{10}
\providecommand{\url}[1]{#1}
\csname url@samestyle\endcsname
\providecommand{\newblock}{\relax}
\providecommand{\bibinfo}[2]{#2}
\providecommand{\BIBentrySTDinterwordspacing}{\spaceskip=0pt\relax}
\providecommand{\BIBentryALTinterwordstretchfactor}{4}
\providecommand{\BIBentryALTinterwordspacing}{\spaceskip=\fontdimen2\font plus
\BIBentryALTinterwordstretchfactor\fontdimen3\font minus
  \fontdimen4\font\relax}
\providecommand{\BIBforeignlanguage}[2]{{%
\expandafter\ifx\csname l@#1\endcsname\relax
\typeout{** WARNING: IEEEtran.bst: No hyphenation pattern has been}%
\typeout{** loaded for the language `#1'. Using the pattern for}%
\typeout{** the default language instead.}%
\else
\language=\csname l@#1\endcsname
\fi
#2}}
\providecommand{\BIBdecl}{\relax}
\BIBdecl

\bibitem{wu2020light_LCNN}
Z.~Wu, R.~K. Das, J.~Yang, and H.~Li, ``Light convolutional neural network with
  feature genuinization for detection of synthetic speech attacks,'' in
  \emph{INTERSPEECH}, 2020.

\bibitem{liu2023leveraging_transformer}
X.~Liu, M.~Liu, L.~Wang, K.~A. Lee, H.~Zhang, and J.~Dang, ``Leveraging
  positional-related local-global dependency for synthetic speech detection,''
  in \emph{ICASSP}, 2023.

\bibitem{jung2022aasist_graph}
J.-w. Jung, H.-S. Heo, H.~Tak, H.-j. Shim, J.~S. Chung, B.-J. Lee, H.-J. Yu,
  and N.~Evans, ``Aasist: Audio anti-spoofing using integrated spectro-temporal
  graph attention networks,'' in \emph{ICASSP}, 2022.

\bibitem{tak2022wav2vec_aasist}
H.~Tak, M.~Todisco, X.~Wang, J.-w. Jung, J.~Yamagishi, and N.~Evans,
  ``Automatic speaker verification spoofing and deepfake detection using
  wav2vec 2.0 and data augmentation,'' in \emph{Proc. The Speaker and Language
  Recognition Workshop (Odyssey)}, 2022.

\bibitem{zhang2024remember_RWM}
X.~Zhang, J.~Yi, C.~Wang, C.~Y. Zhang, S.~Zeng, and J.~Tao, ``What to remember:
  Self-adaptive continual learning for audio deepfake detection,'' in
  \emph{Proceedings of the AAAI Conference on Artificial Intelligence},
  vol.~38, no.~17, 2024, pp. 19\,569--19\,577.

\bibitem{zhang2023you_RAWM}
X.~Zhang, J.~Yi, J.~Tao, C.~Wang, and C.~Y. Zhang, ``Do you remember?
  overcoming catastrophic forgetting for fake audio detection,'' in
  \emph{International Conference on Machine Learning}.\hskip 1em plus 0.5em
  minus 0.4em\relax PMLR, 2023, pp. 41\,819--41\,831.

\bibitem{wang2024comprehensive_CLSurveyForgetting}
Z.~Wang, E.~Yang, L.~Shen, and H.~Huang, ``A comprehensive survey of forgetting
  in deep learning beyond continual learning,'' \emph{IEEE Transactions on
  Pattern Analysis and Machine Intelligence}, 2024.

\bibitem{wang2024comprehensive_CLSurvey}
L.~Wang, X.~Zhang, H.~Su, and J.~Zhu, ``A comprehensive survey of continual
  learning: theory, method and application,'' \emph{IEEE Transactions on
  Pattern Analysis and Machine Intelligence}, 2024.

\bibitem{salvi2025freeze}
D.~Salvi, V.~Negroni, L.~Bondi, P.~Bestagini, and S.~Tubaro, ``Freeze and
  learn: Continual learning with selective freezing for speech deepfake
  detection,'' in \emph{ICASSP}, 2025.

\bibitem{febrinanto2023graph_GLL}
F.~G. Febrinanto, F.~Xia, K.~Moore, C.~Thapa, and C.~Aggarwal, ``Graph lifelong
  learning: A survey,'' \emph{IEEE Computational Intelligence Magazine},
  vol.~18, no.~1, pp. 32--51, 2023.

\bibitem{yu2025select_languageVisionCL}
Y.-C. Yu, C.-P. Huang, J.-J. Chen, K.-P. Chang, Y.-H. Lai, F.-E. Yang, and
  Y.-C.~F. Wang, ``Select and distill: Selective dual-teacher knowledge
  transfer for continual learning on vision-language models,'' in
  \emph{European Conference on Computer Vision}.\hskip 1em plus 0.5em minus
  0.4em\relax Springer, 2025, pp. 219--236.

\bibitem{ma2021continual_DFWF}
H.~Ma, J.~Yi, J.~Tao, Y.~Bai, Z.~Tian, and C.~Wang, ``Continual learning for
  fake audio detection,'' in \emph{INTERSPEECH}, 2021.

\bibitem{chen2025region}
Y.~Chen, J.~Yi, C.~Fan, J.~Tao, Y.~Ren, S.~Zeng, C.~Y. Zhang, X.~Yan, H.~Gu,
  J.~Xue \emph{et~al.}, ``Region-based optimization in continual learning for
  audio deepfake detection,'' in \emph{Proceedings of the AAAI Conference on
  Artificial Intelligence}, vol.~39, no.~22, 2025, pp. 23\,651--23\,659.

\bibitem{chen2025continual}
X.~Chen, W.~Lu, R.~Zhang, J.~Xu, X.~Lu, L.~Zhang, and J.~Wei, ``Continual
  unsupervised domain adaptation for audio deepfake detection,'' in
  \emph{ICASSP}, 2025.

\bibitem{krutsylo2024inter_rehearsalBetter}
A.~Krutsylo, ``The inter-batch diversity of samples in experience replay for
  continual learning,'' in \emph{Proceedings of the AAAI Conference on
  Artificial Intelligence}, vol.~38, no.~21, 2024, pp. 23\,395--23\,396.

\bibitem{rolnick2019experience_ER}
D.~Rolnick, A.~Ahuja, J.~Schwarz, T.~Lillicrap, and G.~Wayne, ``Experience
  replay for continual learning,'' \emph{Advances in neural information
  processing systems}, vol.~32, 2019.

\bibitem{caccia2022new_reservoir_ERACE}
L.~Caccia, R.~Aljundi, N.~Asadi, T.~Tuytelaars, J.~Pineau, and E.~Belilovsky,
  ``New insights on reducing abrupt representation change in online continual
  learning,'' in \emph{International Conference on Learning Representations},
  2022.

\bibitem{hanmo2024effective_rehearsalBetter}
L.~Hanmo, D.~Shimin, L.~Haoyang, L.~Shuangyin, C.~Lei, and Z.~Xiaofang,
  ``Effective data selection and replay for unsupervised continual learning,''
  in \emph{2024 IEEE 40th International Conference on Data Engineering
  (ICDE)}.\hskip 1em plus 0.5em minus 0.4em\relax IEEE, 2024, pp. 1449--1463.

\bibitem{rolnick2019experience_reservoir}
D.~Rolnick, A.~Ahuja, J.~Schwarz, T.~Lillicrap, and G.~Wayne, ``Experience
  replay for continual learning,'' \emph{Advances in neural information
  processing systems}, vol.~32, 2019.

\bibitem{rebuffi2017icarl_herding}
S.-A. Rebuffi, A.~Kolesnikov, G.~Sperl, and C.~H. Lampert, ``icarl: Incremental
  classifier and representation learning,'' in \emph{Proceedings of the IEEE
  conference on Computer Vision and Pattern Recognition}, 2017, pp. 2001--2010.

\bibitem{aljundi2019gradient_gradientSampling}
R.~Aljundi, M.~Lin, B.~Goujaud, and Y.~Bengio, ``Gradient based sample
  selection for online continual learning,'' \emph{Advances in neural
  information processing systems}, vol.~32, 2019.

\bibitem{aljundi2019online_MIR}
R.~Aljundi, E.~Belilovsky, T.~Tuytelaars, L.~Charlin, M.~Caccia, M.~Lin, and
  L.~Page-Caccia, ``Online continual learning with maximal interfered
  retrieval,'' \emph{Advances in neural information processing systems},
  vol.~32, 2019.

\bibitem{chrysakis2020online_classBalance}
A.~Chrysakis and M.-F. Moens, ``Online continual learning from imbalanced
  data,'' in \emph{International Conference on Machine Learning}.\hskip 1em
  plus 0.5em minus 0.4em\relax PMLR, 2020, pp. 1952--1961.

\bibitem{liu2019self_auxiliaryTask}
S.~Liu, A.~Davison, and E.~Johns, ``Self-supervised generalisation with meta
  auxiliary learning,'' \emph{Advances in Neural Information Processing
  Systems}, vol.~32, 2019.

\bibitem{ding2023mitigating_conflictingGradients}
C.~Ding, Z.~Lu, S.~Wang, R.~Cheng, and V.~N. Boddeti, ``Mitigating task
  interference in multi-task learning via explicit task routing with
  non-learnable primitives,'' in \emph{Proceedings of the IEEE/CVF Conference
  on Computer Vision and Pattern Recognition}, 2023, pp. 7756--7765.

\bibitem{baevski2023efficient_maskedObjective}
A.~Baevski, A.~Babu, W.-N. Hsu, and M.~Auli, ``Efficient self-supervised
  learning with contextualized target representations for vision, speech and
  language,'' in \emph{International Conference on Machine Learning}.\hskip 1em
  plus 0.5em minus 0.4em\relax PMLR, 2023, pp. 1416--1429.

\bibitem{todisco2019asvspoof_asvspoof2019}
M.~Todisco, X.~Wang, V.~Vestman, M.~Sahidullah, H.~Delgado, A.~Nautsch,
  J.~Yamagishi, N.~Evans, T.~Kinnunen, and K.~A. Lee, ``Asvspoof 2019: Future
  horizons in spoofed and fake audio detection,'' in \emph{INTERSPEECH}, 2019.

\bibitem{muller2022does_itw}
N.~M. M{\"u}ller, P.~Czempin, F.~Dieckmann, A.~Froghyar, and K.~B{\"o}ttinger,
  ``Does audio deepfake detection generalize?'' in \emph{INTERSPEECH}, 2022.

\bibitem{ma2024cfad}
H.~Ma, J.~Yi, C.~Wang, X.~Yan, J.~Tao, T.~Wang, S.~Wang, and R.~Fu, ``Cfad: A
  chinese dataset for fake audio detection,'' \emph{Speech Communication}, vol.
  164, p. 103122, 2024.

\bibitem{baevski2020wav2vec}
A.~Baevski, Y.~Zhou, A.~Mohamed, and M.~Auli, ``wav2vec 2.0: A framework for
  self-supervised learning of speech representations,'' \emph{NeurIPS}, 2020.

\bibitem{jung2022aasist_aasist}
J.-w. Jung, H.-S. Heo, H.~Tak, H.-j. Shim, J.~S. Chung, B.-J. Lee, H.-J. Yu,
  and N.~Evans, ``Aasist: Audio anti-spoofing using integrated spectro-temporal
  graph attention networks,'' in \emph{ICASSP}, 2022.

\bibitem{kirkpatrick2017overcoming_EWC}
J.~Kirkpatrick, R.~Pascanu, N.~Rabinowitz, J.~Veness, G.~Desjardins, A.~A.
  Rusu, K.~Milan, J.~Quan, T.~Ramalho, A.~Grabska-Barwinska \emph{et~al.},
  ``Overcoming catastrophic forgetting in neural networks,'' \emph{Proceedings
  of the national academy of sciences}, vol. 114, no.~13, pp. 3521--3526, 2017.

\bibitem{li2017learning_LWF}
Z.~Li and D.~Hoiem, ``Learning without forgetting,'' \emph{IEEE transactions on
  pattern analysis and machine intelligence}, vol.~40, no.~12, pp. 2935--2947,
  2017.

\bibitem{zeng2019continual_OWM}
G.~Zeng, Y.~Chen, B.~Cui, and S.~Yu, ``Continual learning of context-dependent
  processing in neural networks,'' \emph{Nature Machine Intelligence}, vol.~1,
  no.~8, pp. 364--372, 2019.

\end{thebibliography}

\end{document}